\documentclass[tpreprintnumbers,amsmath,amssymb,nofootinbib,longbibliograpphy]{revtex4}

\usepackage{dcolumn}
\usepackage{mathrsfs}
\usepackage{amsmath}
\usepackage{graphicx}
\usepackage{bm}
\usepackage{epstopdf}
\usepackage{float}
\usepackage{hyperref}
\usepackage{color}

\usepackage{amssymb}
\usepackage{eucal}
\usepackage{mathrsfs}
\usepackage{amsthm}

\usepackage{changes}

\def\bk{{\bf k}}
\def\bkp{{\bf k}'}
\def\wk{{\omega_k}}
\def\wkp{{\omega_{k'}}}
\def\akl{a_{\bk \lambda}}
\def\aklp{a_{\bk ' \lambda '}}
\def\akld{a_{\bk \lambda}^{\dagger}}
\def\aklpd{a_{\bk '\lambda '}^{\dagger}}
\def\bfkl{{\bf f}_{\bk \lambda}}
\def\bfkls{{\bf f}_{\bk \lambda}^*}
\def\bfklp{{\bf f}_{\bk ' \lambda '}}
\def\bfklps{{\bf f}_{\bk ' \lambda '}^*}
\def\br{{\bf r}}

\def\bE{{\bf E}}
\def\ekr{e^{i\bk \cdot \br}}

\def\ekl{{\hat{{\bf e}}}_{\bk \lambda}}
\def\eklp{{\hat{{\bf e}}}_{\bkp \lambda '}}
\def\bq{{\bf q}}
\def\b0{{\bf 0}}
\def\vac{{\{ 0_{\bk \lambda} \}}}
\def\bR{{\bf R}}

\begin{document}

\title{Effective Hamiltonians in Nonrelativistic Quantum~Electrodynamics}

\author{Roberto Passante$^{1,2}$\footnote{roberto.passante@unipa.it}}
\author{Lucia Rizzuto$^{1,2}$\footnote{lucia.rizzuto@unipa.it}}

\affiliation{$^1$ Dipartimento di Fisica e Chimica - Emilio Segr\`{e},
Universit\`{a} degli Studi di Palermo, Via Archirafi 36, I-90123 Palermo,
Italy}
\affiliation{$^2$ INFN, Laboratori Nazionali del Sud, I-95123 Catania, Italy}

\begin{abstract}
In this paper, we consider some {second-order} effective Hamiltonians describing the interaction of the quantum electromagnetic field with atoms or molecules in the nonrelativistic limit.
{Our procedure is valid only for off-energy-shell processes, specifically virtual processes such as those relevant for ground-state energy shifts and dispersion van der Waals and Casimir--Polder interactions, while on-energy-shell processes are excluded.}
 These effective Hamiltonians allow for a considerable simplification of the calculation of radiative energy shifts, dispersion, and Casimir--Polder interactions, including in the presence of boundary conditions. They can also provide clear physical insights into the processes involved. We clarify that the form of the effective Hamiltonian depends on the field states considered, and consequently different expressions can be obtained, each of them with a well-defined range of validity and possible applications.  We also apply our results to some specific cases, mainly the Lamb shift, the Casimir--Polder atom-surface interaction, and the dispersion interactions between atoms, molecules, or, in general, polarizable bodies.
 \end{abstract}
 
 \maketitle

\section{Introduction}
\label{Introduction}

In molecular quantum electrodynamics, that is, the quantum theory of atoms and molecules interacting with the electromagnetic field in the nonrelativistic limit, several processes of great interest are of a high order in atom--field coupling~\cite{Salam10,Salam08,Milonni19}.
For example, interatomic dispersion interactions such as van der Waals and Casimir--Polder interactions between two atoms or molecules are fourth-order processes~\cite{Casimir--Polder48,Compagno-Passante-Persico95,Passante18}, and, in~the case of three or more atoms, many-body effects start from the sixth order~\cite{Aub-Zienau60,Salam10,Aldegunde-Salam15}. Additionally, the resonance energy transfer between molecules may involve high-order perturbative calculations in the atom--field coupling~\cite{Salam21}. In~such cases, the~number of relevant Feynman diagrams rapidly grows with the perturbative order, with~consequent increasing complexity of the calculations. For~this reason, the possibility of finding approximated effective Hamiltonians allowing for the simplification of calculations is very important, and, hopefully, this will also yield a transparent interpretation and physical insights into the relevant physical processes involved~\cite{Passante-Power98,Barcellona-Safari17,Passante18,Buhmann-Salam18,Craig-Power69,Passante-Power87}. Effective Hamiltonians can be also used in dynamical (time-dependent) nonequilibrium situations~\cite{Rizzuto-Passante04}. All these possibilities have fostered the investigation of effective Hamiltonians containing an interaction term that is at least quadratic in the atom--field coupling,
where the response of the atom is included in quantities such as, for~example, its polarizability,
thus allowing for a considerable reduction of the perturbative order required for calculating specific processes and of the number of relevant Feynman diagrams~\cite{Passante18}.
Very recently, resummation techniques, in~which the polarizability is summed to any order, have been developed, and~they could also be of great importance for the evaluation of radiative processes such as the Lamb shift and van der Waals interactions for nanostructured materials~\cite{Flick-Schafer18,Haugland-Schafer21,Schafer-Buchholz21}.

In this paper, we obtain and review different forms of effective Hamiltonians, {at the second order in the atom--field coupling,} used in nonrelativistic quantum electrodynamics, stressing the range and limit of validity of each of them, as~well as their physical interpretation. Application to retarded and nonretarded dispersion interactions and to the atom-surface Casimir--Polder interaction is also discussed. A~physical interpretation in terms of the response of the atoms or molecules to vacuum field fluctuations or to real photons is also outlined, as~well as the possible role of dissipation in the response function.
{More specifically, we here present in a clear and organic way the different forms of second-order effective Hamiltonians that can be obtained, depending on the specific field states considered. We also show and stress that the response of the atom is different in the case of vacuum fluctuations or when photons are present; in fact, we find that in the first case, the atom responds through a real function of the frequency, while in the second case through its dynamical polarizability, that has poles in the real frequency axis.}

This paper is organized as follows. In~Section~\ref{EffHam}, the~general expression of the effective Hamiltonian and of its matrix elements between atom+field states is obtained. In \mbox{Section~\ref{EH}}, specific forms of the effective Hamiltonian in relevant cases are given, and~their limits of validity are stressed, as~well as the response function of the atoms in the various cases considered, and~the possible role of dissipation. Application to interatomic dispersion interactions and to the Casimir--Polder atom-surface interaction, both in the nonretarded and retarded cases, is outlined. Finally, Section~\ref{Conclusions} is devoted to our conclusive~remarks.

\section{The General Expression of the Effective Hamiltonian in Molecular Quantum~Electrodynamics}
\label{EffHam}

We first start from the Hamiltonian of one atom, placed at
$\bR$, interacting with the quantum electromagnetic field, in~the multipolar coupling scheme and within dipole approximation~\cite{Compagno-Passante-Persico95}. Extension to the case of two or more atoms is straightforward, at~the order considered. The~Hamiltonian of the system is
\vspace{-15pt}

\begin{eqnarray}
\label{Hamiltonian}
H &=& H_0 + H_I , \nonumber \\
H_0 &=& \sum_{\bk \lambda} \hbar \wk \akld \akl + \sum_\ell E_\ell \lvert \phi_\ell \rangle \langle \phi_\ell \rvert ,  \nonumber \\
H_I &=&  - {\bm \mu} \cdot \bE (\bR ) = -\sum_{\ell m}e{\bq}^{\ell m}\cdot \bE (\bR ) \lvert \phi_\ell \rangle \langle \phi_m \rvert ,
\end{eqnarray}
where states $\lvert \phi_\ell \rangle$ are a complete set of atomic states with energies $E_\ell$, and $\akld$ and $\akl$ are, respectively, creation and annihilation operators relative to the field mode $(\bk \lambda )$, with $\lambda =1,2$ being the polarization index and~satisfying the usual bosonic commutation relations, and~$\wk =ck$; ${\bm \mu} = e\bq$ is the electric dipole moment operator of the atom, with $\bq$ being the electron coordinate.
For simplicity, we assume only one atomic electron is taking part in the radiative process considered, and~${\bq}^{\ell m} = \langle \phi_\ell \lvert \bq \rvert \phi_m \rangle$ is its matrix element between atomic states.
{Although this simplification, which is strictly valid for hydrogen-like systems, is not essential in the present calculation, it allows one to include one-electron dipole moment matrix elements only and simplify the discussion of the results; it can, however, be simply overtaken by introducing a full dipole moment operator.}

 Finally, $\bE (\br )$ is the electric field operator
\begin{equation}
\label{ElectricField}
\bE (\br ) = \sum_{\bk \lambda} \bE (\bk \lambda ; \br ) = \sum_{\bk \lambda} \left( \bfkl (\br ) \akl + \bfkl^* (\br ) \akld \right) ,
\end{equation}
where $\bE (\bk \lambda ; \br )$ is a Fourier component of the electric field, with~$\bfkl (\br )$ the field mode functions taking into account the boundary conditions present. In~the free space, we have
\begin{equation}
\label{FreeSpace}
\bfkl (r) = i\sqrt{\frac {2\pi \hbar \wk}V} \ekl \ekr ,
\end{equation}
where $\ekl$ $(\lambda =1,2)$ are polarization unit vectors, such that $\ekl \cdot \hat{{\bf e}}_{\bk \lambda '} = \delta_{\lambda \lambda '}$, $\ekl \cdot \hat{\bk}=0$,  and~$V$ is the quantization~volume.

The multipolar-coupling Hamiltonian (\ref{Hamiltonian}) contains also a term equal to $2\pi \int d^3r ({\bf P}_\perp (\br ))^2$, where ${\bf P}_\perp (\br )$ is the transverse part of the polarization field
${\bf P}(\br )= \sum_i e\br \delta(\br -\br_i)$, with $\br_i$ being the position of an atomic electron~\cite{Power64,Milonni76,Salam10}. This is a second-order term in the electric charge that, although~it is important for the Lamb shift, does not contribute to the dispersion interactions~\cite{Milonni76,Craig-Thirunamachandran98,Salam10,Passante18}, and~for this reason we do not include it, except~whenever necessary.

We now follow and briefly review the standard general procedure used in nonrelativistic quantum electrodynamics to obtain a second-order effective Hamiltonian, which consists of applying the following transformation to the Hamiltonian (\ref{Hamiltonian})
\begin{equation}
\label{Transformation}
T = \exp{\left( iez/\hbar \right)} ,
\end{equation}
where $z$ is an Hermitian operator (with some limitations on the state space where it is defined, as~specified later on), chosen in such a way to eliminate, in~the transformed Hamiltonian, the~first-order terms in the electron charge $e$ \cite{Passante-Power98,Passante-Power87,Craig-Power69}. Up~to the second order in $e$, we have
\begin{eqnarray}
\label{Transformation1}
T^{-1}HT &=& \exp{\left( -iez/\hbar \right)}H\exp{\left( iez/\hbar \right)}
= H_0 -e \bq \cdot \bE (\bR )+ \left[ -iez/\hbar ,H_0 \right]
\nonumber \\
&\ & + \left[  -iez/\hbar , -e \bq \cdot \bE (\bR )\right] + \frac 12 \left[ -iez/\hbar , \left[ -iez/\hbar ,H_0 \right]\right] + O(e^3) .
\end{eqnarray}

The operator $z$ is chosen in such a way to make vanishing the first-order terms in the transformed Hamiltonian (\ref{Transformation1}), that is
\begin{equation}
\label{Condition}
-e \bq \cdot \bE (\bR ) -\frac {ie}\hbar \left[ z ,H_0 \right] =0 ,
\end{equation}
so that the transformed Hamiltonian becomes
\begin{equation}
\label{TransfHam}
\tilde{H} = T^{-1}HT = H_0 + \frac {ie^2}{2\hbar} \left[ z, \bq \cdot \bE (\bR ) \right]
= H_0 + {\tilde{H}}_{eff},
\end{equation}
with the effective Hamiltonian, expressed in terms of the Hermitian operator $z$, given by
\begin{equation}
\label{EffectiveH}
{\tilde{H}}_{eff} = \frac {ie^2}{2\hbar} \left[ z, \bq \cdot \bE (\bR ) \right] .
\end{equation}

This Hamiltonian is at the second order in the atom--field coupling. A~second-order energy shift of the system due to the atom--field coupling is then given by the average value of ${\tilde{H}}_{eff}$ on the state of the system at hand. Fourth-order corrections, as~in the evaluation of interatomic dispersion interactions, can be obtained by a second-order approach if the effective Hamiltonian is used, as~explicitly discussed in the next~section.

Equation~(\ref{Condition}) is an implicit relation defining the operator $z$ of the transformation. From~this relation, we can obtain its matrix elements between atom+field states of the generic form $\lvert \{p\}, \phi_P \rangle$, $\{p\}$, and $\phi_P$, denoting, respectively, generic field and atom states ($\{p\}$ denotes a general set of number states of the field for all the modes allowed by the boundary conditions), with~energy ${E_p=} \sum_i \hbar \omega_i p_i$ and $E_P$, respectively. We obtain
\begin{eqnarray}
\label{MatrixEl1}
&\ & \langle\{ m\}, \phi_M \rvert \bq \cdot \bE (\bR ) \lvert \{n\}, \phi_N \rangle \nonumber \\
&\ & =-\frac i\hbar  \langle \{m\}, \phi_M \rvert z \lvert \{n\}, \phi_N \rangle
\left( E_n -E_m + E_N -E_M \right)
\end{eqnarray}

Assuming $E_n -E_m + E_N -E_M \neq 0$, that is, that the matrix element is taken between unperturbed states with a different energy, we get the matrix elements of the operator $z$
\begin{equation}
\label{MatrixEl2}
\langle \{m\}, \phi_M \rvert z \lvert \{n\}, \phi_N \rangle =i\hbar
\frac {\langle \{m\}, \phi_M \rvert \bq \cdot \bE (\bR ) \lvert \{n\}, \phi_N \rangle}{(E_n+E_N)-(E_m+E_M)}
\end{equation}

Equation~(\ref{MatrixEl2}) defines all matrix elements of $z$, except~those on the energy shell, that are undefined; thus, the~transformation (\ref{Transformation}) with the matrix elements (\ref{MatrixEl2}) is not defined on the energy shell. Therefore, the~effective Hamiltonian so obtained is not valid when energy-conserving processes are relevant, unless~some regularization procedure of the vanishing energy denominators, taking into account dissipation, is used~\cite{Andrews-Naguleswaran98,Milonni-Boyd04,Milonni-Loudon08}.
This regularization can be done by adding an imaginary part to the energy denominators shifting the poles to the lower complex frequency half-plane, in~agreement with the causality requirement; in our case, this can be done only phenomenologically because we are working within a second-order perturbative approach for a closed system.
Because we will obtain effective Hamiltonians mainly to calculate atomic energy shifts and (many-body) dispersion interactions for ground-state systems, where only virtual processes are involved, or~when resonant processes are suppressed (in a cavity, for~example), in~general, this is not a strong limitation for our purposes. However, we wish to point out that our results cannot be directly extended to the case of excited atomic states when resonant processes are~present.

For the reasons mentioned above, substitution of the matrix elements (\ref{MatrixEl2}) of $z$ into (\ref{Transformation1}) allows us to obtain only the off-energy-shell matrix elements of the effective Hamiltonian and~not its full expression as an operator in the Hilbert space of the system.
{For this reason, we can obtain different expressions in terms of field operators, according to the type of field states at hand.}
The matrix elements of the effective Hamiltonian are
\begin{eqnarray}
\label{EffHamMatrixElements}
&\ & \langle \{m\}, \phi_M \rvert {\tilde{H}}_{eff} \lvert \{n\}, \phi_N \rangle
\nonumber \\
 &\ & = -\frac 12 e^2 \sum_{\{\ell\} L} \langle \{m\}, \phi_M \rvert  \bq \cdot \bE (\bR ) \lvert \{\ell\} , \phi_L \rangle \langle
\{\ell\} , \phi_L \rvert \bq \cdot \bE (\bR ) \lvert \{n\}, \phi_N \rangle
\nonumber \\
&\ & \times \left( \frac 1{E_L -E_M + \sum_s \left( \ell_s -m_s \right) \hbar \omega_s} - \frac 1{E_N -E_L + \sum_s \left( n_s -\ell_s \right) \hbar \omega_s}
\right) ,
\end{eqnarray}
where the index $s$ indicates photonic modes with frequency $\omega_s = c k_s$, and~the sum over $s$ runs over all field modes. Using the mode expansion (\ref{ElectricField}) for the electric field operator, the~matrix elements appearing in the RHS of (\ref{EffHamMatrixElements}) can be cast in the form
\begin{eqnarray}
\label{MatrElem}
&\ &\langle \{m\}, \phi_M \rvert  \bq \cdot \bE (\bR ) \lvert \{\ell\} , \phi_L \rangle \langle \{\ell\} , \phi_L \rvert \bq \cdot \bE (\bR ) \lvert \{n\}, \phi_N \rangle
\nonumber \\
&=& \sum_L \sum_ {\bk \lambda \bkp \lambda '} \langle \{m\}, \phi_M \rvert  \bq \cdot \left( \bfklp (\bR ) \aklp + \bfklps (\bR ) \aklpd \right) \lvert \{\ell\} , \phi_L \rangle
\nonumber \\
&\ & \ \times \langle \{l\}, \phi_L \rvert  \bq \cdot \left( \bfkl (\bR ) \akl + \bfkls (\bR ) \akld \right) \lvert \{n\} , \phi_N \rangle .
\end{eqnarray}

From Equation~(\ref{MatrElem}), it is easy to see that, in~order that the quantity above be nonvanishing, it is necessary that the total number of photons in $\{m\}$ and $\{n\}$ (that is, the photon number summed over all field modes) must be equal or differ by two. There are thus three possibilities: (i) $\{n\} = \{m\}$, that is elements diagonal in the photon space state; (ii) the difference in $\{n\}$ and $\{m\}$ is all in a single mode
$(\bar{\bk} \bar{\lambda})$, and~thus $m_{\bar{\bk} \bar{\lambda}} = n_{\bar{\bk} \bar{\lambda}} \pm 2$, with all other modes containing the same number of photons; (iii) the difference is by one photon in each of the two modes $(\bar{\bk} \bar{\lambda})$ and $(\tilde{\bk} \tilde{\lambda})$, with $(\bar{\bk} \bar{\lambda}) = n_{\bar{\bk} \bar{\lambda}} \pm 1$ and $(\tilde{\bk} \tilde{\lambda}) = n_{\tilde{\bk} \tilde{\lambda}} \pm 1$ (upper or lower sign for both modes), while all other modes in $\{m\}$ and $\{n\}$ have the same photon~number.

In the next section, we will explicitly find the form of the effective Hamiltonian for some specific cases, which is relevant in molecular quantum electrodynamics, in~particular for atom-surface and (many-body) atom--atom interactions for ground-state atoms or molecules. On~the basis of the results obtained in this section, we will now find the relative effective Hamiltonians in the various cases, according to the relevant photon states involved, and~point out the range and limit of application of the specific forms~obtained.

\section{Diagonal and Off-Diagonal Matrix Elements of the Effective~Hamiltonian}
\label{EH}

We now evaluate the matrix elements of the effective Hamiltonian introduced in the previous section. As~mentioned in the previous section, the~expression of the effective Hamiltonian operator that we obtain, which acts only in the field space, and~of its matrix elements, can differ according to the subspace of photon states considered. We work in the Schr\"{o}dinger representation, and~the energy shifts are to be evaluated by time-independent perturbation theory. Related effective Hamiltonians for evaluating energy shifts and dispersion interactions have been also used with a different approach, specifically in the Heisenberg representation in terms of time-dependent field operators, and~by separating free and source (scattering) fields~\cite{Power-Thirunamachandran83,Shahmoon15,Milonni94,Milonni-Smith96,Milonni07}.

We now separately consider different cases; we~obtain the relative explicit expressions of the effective Hamiltonian operator (off the energy shell), and~apply them to some relevant physical systems, mainly van der Waals and Casimir--Polder dispersion interactions. The~possible role of dissipation will be also discussed. The~results we obtain are relative to a microscopic description of the system; however, we argue that some aspects of our results could be highly relevant also for macroscopic polarizable bodies and Casimir interactions between macroscopic dielectric bodies in the~vacuum.

\subsection{Diagonal Elements of the Effective Hamiltonian and Ground-State Systems (Zero Photons)}
\label{GroundState}

Let us first consider the unperturbed ground state of the system, that is, $\lvert \vac , g \rangle$, where $\lvert \vac \rangle$ denotes the photon vacuum state and $\lvert g \rangle$ is the atom's ground state. Due to the atom--field coupling, this state is not an eigenstate of the interacting Hamiltonian, and, as~it is well known, the~interaction leads to a second-order energy shift that, after~mass renormalization, yields the nonrelativistic Lamb shift of the ground-state atom~\cite{Bethe47,Power64,Compagno-Passante-Persico95,Maclay20}. This energy shift is a second-order effect in the electron~charge.

Using our effective Hamiltonian (\ref{EffHamMatrixElements}), this energy shift can be obtained just through a simple first-order calculation, namely, as the expectation value of the effective Hamiltonian on the unperturbed state~\cite{Passante-Power87}. In~fact, from~(\ref{EffHamMatrixElements}), and~using (\ref{ElectricField}), we immediately obtain (Einstein's convention of summation over the repeated indices $i,j$ is assumed)
\begin{eqnarray}
\label{EnergyShift1}
&\ & \langle \vac ,g \rvert {\tilde{H}}_{eff} \lvert \vac ,g \rangle
\nonumber \\
&\ & = - \frac 12 e^2\sum_{\bk \lambda} \sum_L
\frac {2\langle \vac ,g \rvert \bq \cdot \bE (\bR )\lvert 1_{\bk \lambda} ,\phi_L \rangle \langle 1_{\bk \lambda} , \phi_L \rvert \bq \cdot \bE (\bR )\lvert \vac ,g \rangle }{E_{L g}+\hbar \wk}
\nonumber \\
&\ & = -\frac 12 \sum_{\bk \lambda} \left( \sum_L \frac {2\mu_i^{gL} \mu_j^{Lg}}{E_{Lg}+\hbar \wk} \right) \langle \vac \rvert E_i(\bk \lambda ; \bR ) E_j(\bk \lambda ; \bR )\lvert \vac \rangle ,
\end{eqnarray}
where $\lvert 1_{\bk \lambda} \rangle$ are one-photon states, $\lvert \phi_L \rangle$ are intermediate atomic states, and~$E_{L g} = E_L -E_g$; also, $E_{L g} > 0$, and~therefore the quantity in the round brackets in (\ref{EnergyShift1}) has no poles.
{It involves off-energy-shell processes only.}
 Thus, in~(\ref{EnergyShift1}) there is no need to introduce dissipation, and, very importantly, the~real quantity
\begin{equation}
\label{grstatepol}
\beta_{ij}^g (\wk ) = \sum_L \frac {2\mu_i^{gL} \mu_j^{Lg}}{E_{Lg}+\hbar \wk}
\end{equation}
appears, ~not the atomic dynamic polarizability (see also the following discussion). All relevant properties of the atom are embedded in this function. Equation~(\ref{EnergyShift1}) is the average value on the vacuum state of the electromagnetic field $\vac$ of the following field operator
\begin{equation}
\label{EffHam0}
H_{eff}^0 = -\frac 12 \sum_{\bk \lambda} \beta_{ij}^g(\wk ) E_i(\bk \lambda ; \bR ) E_j (\bk \lambda ;\bR ),
\end{equation}
that we can take as our effective Hamiltonian within the zero-photon subspace, as~specified by the apex $0$.

By taking the average value of  (\ref{EffHam0}) on the field vacuum state $\lvert \vac \rangle$, we immediately obtain the second-order energy shift due to the atom--field interaction
\begin{equation}
\label{EnergyShift0}
\Delta E_g = \langle \vac \lvert H_{eff}^0 \rvert \vac \rangle = -\frac 12 \sum_{\bk \lambda} \beta_{ij}^g (\wk ) [ \bfkls (\bR )]_i [ \bfkl (\bR )]_j  ,
\end{equation}
where $\bfkl (\bR )$ are the appropriate mode functions for the system at hand, as~introduced in Equation~(\ref{ElectricField}), ~evaluated at the position $\bR$ of the atom. For~example, in~the case of an atom in the unbounded space, Equation~(\ref{EnergyShift0}) yields its ground-state Lamb shift~\cite{Passante-Power87}
{(as previously mentioned, the~second-order term proportional to ${\bf P}_\perp^2 (\br )$ should be added in this case),}
while for an atom in front of a reflecting plate it yields the atom--surface Casimir--Polder interaction energy at zero temperature~\cite{Messina-Passante08}.

The effective Hamiltonian (\ref{EffHam0}) has a clear physical interpretation. Vacuum electric field fluctuations with wavevector $\bk$ induce a dipole moment on the atom given by
\begin{equation}
\label{InducedDipole}
(\mu_{ind}^0)_i(\bk ) \sim \beta_{ij}^g(\wk ) E_j(\bk \lambda ; \bR ) ,
\end{equation}
and, in~turn, this fluctuating dipole moment interacts with the vacuum field fluctuations yielding (we assume that a vacuum expectation value is taken)
\begin{equation}
\label{IntEn}
\Delta E \sim -\frac 12 \sum_{\bk \lambda} (\mu_{ind}^0)_i(\bk ) E_i(\bk \lambda ; \bR ),
\end{equation}
finally yielding (\ref{EffHam0}) and (\ref{EnergyShift0}). In~this case, it is as if a Fourier component of the induced dipole moment with a given $\bk$ interacts only with the Fourier component of the electric field having the same $\bk$. As~we will show in the next subsection, this does not occur when off-diagonal elements of the effective Hamiltonian are relevant, as~in the case of two- and many-body dispersion interactions between atoms or~molecules.

A very important point is that the response of the atom to the vacuum field fluctuations is not through its dynamical polarizability but~through the function $\beta_{ij}^g (\wk )$, as~defined by (\ref{grstatepol}), which is a real quantity at any frequency without poles and related dissipative properties.
This does not contrast with the fluctuation-dissipation theorem and the Kramers--Kr\"{o}nig dispersion relations, which indeed consider the response of a system to an external applied field (linear response theory), while in our case only vacuum field fluctuations act, which cannot induce real transitions. This result could be important also for macroscopic polarizable bodies and their role as boundary conditions in the Casimir effect, as~well as for the long-lasting dispute in the literature about the most appropriate dielectric model (plasma or Drude model, for~example) to be used in the Casimir effect for dielectrics~\cite{Bordag-Klimchitskaya09,Mostepanenko21,Brevik-Shapiro21}. We will address this point in a forthcoming~publication.

From Equations~(\ref{EffHamMatrixElements}) and (\ref{MatrElem}), we can also obtain diagonal matrix elements of the effective Hamiltonian in the case of one mode populated with $n_{\bk \lambda}$ photons, with all other field modes being in their vacuum state, obtaining
\begin{eqnarray}
\label{EnergyShift_n}
&\ & \langle n_{\bk \lambda} ,\phi_N \rvert {\tilde{H}}_{eff} \lvert n_{\bk \lambda} ,\phi_N \rangle
\nonumber \\
&\ &= -\frac 12 \sum_L \lvert {\bm \mu}^{NL} \cdot \bE (\bk \lambda ; \bR ) \rvert^2
\left( 2n_{\bk \lambda} \frac {2E_{LN}}{E_{LN}^2- (\hbar \wk )^2} + \frac 2{E_{LN}+\hbar \wk} \right)
\nonumber \\
&\ & = -\frac 12 E_i(\bk \lambda ; \bR ) E_j (\bk \lambda ;\bR ) \left( 2n_{\bk \lambda} \alpha_{ij}(\wk ) + \beta_{ij}(\wk ) \right) ,
\end{eqnarray}
where
\begin{equation}
\label{DynPol}
\alpha_{ij}(\wk ) = \sum_L \frac {2E_{LN}\mu^{NL}_i \mu^{LN}_j}{E_{LN}^2- (\hbar \wk )^2}
\end{equation}
is the ground-state atomic dynamic polarizability, with~poles at $\hbar \wk = E_{LN}$ and obeying the dispersion relations, and~\begin{equation}
\label{betaN}
\beta_{ij}^N (\wk ) = \sum_L \frac {2\mu_i^{NL} \mu_j^{LN}}{E_{LN}+\hbar \wk}
\end{equation}
is a function analogous to (\ref{grstatepol}) for the generic atomic state $N$. Equation~(\ref{betaN}) has no poles if $N$ is the ground state, as~already discussed, while it can have poles in the case of other atomic states. We wish to point out and stress that
Equation~(\ref{EnergyShift_n}) clearly shows that the atom responds to real photons through its dynamical polarizability $\alpha_{ij}(\wk )$ and to the vacuum fluctuating field through the $\beta_{ij}^N (\wk )$ function.

For atoms or molecules with a random orientation,
being $\langle \mu_i^{gL} \mu_j^{Lg} \rangle = \mid  {\bm \mu}^{gL}  \mid^2 \delta_{ij}/3$, Equation~(\ref{EffHam0}) becomes
\begin{equation}
\label{EnergyShift3}
H_{eff}^{(0)} =  - \frac 12 \sum_{\bk \lambda} \beta_{av}^g (\wk ) \left( E(\bk \lambda ;\bR )  \right)^2,
\end{equation}
with
\begin{equation}
\label{polgsav}
\beta_{av}^g (\wk ) = \frac 23 \sum_L \frac { \mid  {\bm \mu}^{Lg}  \mid^2}{E_{Lg}+\hbar \wk} .
\end{equation}

If the relevant frequencies of the field are such that $\wk \ll E_{Lg}/\hbar$, then $\beta_{av}^g (\wk ) \simeq \beta_{av}^g (0)=\alpha^g$, where $\alpha^g$ is the isotropic static polarizability of the ground-state atom. Then, Equation~(\ref{EnergyShift3}) reduces to $H_{eff}^{(0)} \simeq -\frac 12 \alpha E^2(\bR )$; we wish to stress that only in this limiting case, $\wk \rightarrow 0$, our $\beta$ function coincides with the (static) polarizability of the atom. This form of the effective Hamiltonian has been used, for example, for evaluating the far-zone (retarded) dispersion interaction of an atom placed near a conducting wall~\cite{Milonni94}.

\subsection{Off-Diagonal Elements of the Effective~Hamiltonian}
\label{OffDiagElem}

We now consider the off-diagonal elements of the effective Hamiltonian between atom--field states, which we can obtain from (\ref{EffHamMatrixElements}) and (\ref{MatrElem}); as mentioned at the end of the previous section, in~this case the total number of photons in the two states must differ by two. We find
\begin{eqnarray}
\label{OffDiaMatrElem}
&\ & \langle \{ p \} , n_{\bk \lambda}+1, m_{\bkp \lambda '}+1, \phi_M \lvert \tilde{H}_{eff} \rvert \{ p \} , n_{\bk \lambda}, m_{\bkp \lambda '}, \phi_N \rangle
\nonumber \\
&=&  -\frac 12 e^2 \sum_L  \langle \phi_M \lvert q_i \rvert \phi_L \rangle \langle \phi_L \lvert q_j \rvert \phi_N \rangle \sqrt{n_{\bk \lambda}+1} \sqrt{m_{\bkp \lambda '}+1}
(\bfkls (\bR ))_i (\bfklps (\bR ))_j
\nonumber \\
&\ & \times \left( \frac 1{E_{LM} + \hbar \wk} -  \frac 1{E_{NL} + \hbar \wkp} \right) + \left( \bk \lambda \leftrightarrow \bkp \lambda ' \right) ,
\end{eqnarray}
where $\{ p \}$ indicates all modes different from $\bk \lambda$ and $\bkp \lambda '$, $\bk \lambda \neq \bkp \lambda '$.

We can specialize Equation~(\ref{OffDiaMatrElem}) to the case $n_{\bk \lambda} = m_{\bkp \lambda '}=0$, $\{ p \} = \{ 0 \}$ (vacuum state for all other field modes), $\phi_M = \phi_N$, relevant for the calculation of two- or three-body dispersion interactions between atoms or molecules in the vacuum space at zero temperature, allowing one to reduce, respectively, a~fourth or sixth-order calculation to a second- or third-order calculation~\cite{Passante-Power98},
\begin{eqnarray}
\label{OffDiaMatrElem2}
&\ & \langle \{ 0 \} , 1_{\bk \lambda}, 1_{\bkp \lambda '}, \phi_N \lvert \tilde{H}_{eff} \rvert \{ 0 \} , 0_{\bk \lambda}, 0_{\bkp \lambda '}, \phi_N \rangle
\nonumber \\
&=&  -\frac 12 (\bfkls (\bR ))_i (\bfklps (\bR ))_j
\sum_L  \left( \frac {2E_{LN}\mu_i^{NL}\mu_j^{LN} }{E_{LN}^2- (\hbar \wk )^2} + \frac {2E_{LN}\mu_i^{NL}\mu_j^{LN}}{E_{LN}^2- (\hbar \wkp )^2}  \right)
\nonumber \\
&=& -\frac 12 (\bfkls (\bR ))_i (\bfklps (\bR ))_j  \left(  \alpha_{ij} (\wk ) + \alpha_{ij} (\wkp ) \right),
\end{eqnarray}
where the symmetry between primed and non primed $(\bk \lambda )$s has been exploited, and~the atomic matrix elements ${\bm \mu}^{LN}$ can be taken real without loss of generality. This quantity is the matrix element on the field states considered,
$\lvert \{ 0 \} , 0_{\bk \lambda}, 0_{\bkp \lambda '}, \phi_N \rangle$ and $\lvert \{ 0 \} , 1_{\bk \lambda}, 1_{\bkp \lambda '}, \phi_N \rangle$
 of the following operator
\begin{equation}
\label{EffHam2}
H'_{eff} = -\frac 12 \sum_{\bk \lambda} \alpha_{ij} (\wk ) E_i (\bk \lambda ; \bR ) E_j (\bR ) ,
\end{equation}
that we can take as the effective Hamiltonian in the field subspace spanned by the states considered. The~properties of the atoms or molecules relevant for the processes involved are contained in their dynamical polarizability, $\alpha_{ij}(\wk )$, and~the effective Hamiltonian acts on the photon subspace~only.

The effective Hamiltonian (\ref{EffHam2}) is different from (\ref{EffHam0}), and~ its physical interpretation is different. The~response of the atom in the present case is not through the $\beta_{ij}(\wk )$ function but through its dynamical polarizability, and~the interaction of the $\bk$ Fourier component of the induced dipole moment is with the total electric field and~not with only its $\bk$ component, as~in the previous diagonal case (\ref{IntEn}).

If the frequency of the (virtual) photons involved is much smaller than the relevant atomic transition frequencies, we can approximate the dynamical polarizabiblity with the static one, $\alpha_{ij} (\wk )\simeq \alpha_{ij} (0)=\alpha_{ij}$, and~(\ref{EffHam2}) reduce it to
\begin{equation}
\label{EffHamst}
H_{eff}^{st} = -\frac 12 \sum_{\bk \lambda} \alpha_{ij} E_i (\bk \lambda ; \bR ) E_j (\bR ) = -\frac 12 \alpha_{ij} E_i (\bR ) E_j (\bR ) ,
\end{equation}
which can be used to obtain retarded far-zone dispersion interactions, where only the contribution of low-frequency virtual photons is relevant~\cite{Craig-Power69}.

A straightworward application of the effective Hamiltonian (\ref{EffHam2}) is the evaluation of dispersion (van der Waals and Casimir--Polder) interactions between two ground-state atoms in the vacuum (zero temperature), even when boundary conditions are present. In~this case, we have two atoms, labeled as A and B, respectively, located at $\bR_A$ and $\bR_b$. The~standard quantum electrodynamical calculation of their dispersion interaction involves a fourth-order perturbative calculation~\cite{Casimir--Polder48,Craig-Thirunamachandran98}, while the use of the effective Hamiltonian \mbox{(\ref{EffHam2})} allows us to reduce it to a much simpler second-order calculation. The~Hamiltonian of the system is
$H=H_A+H_B+H_F+H'_{eff}(A)+H'_{eff}(B)$, where $H_A$ and $H_B$ are, respectively, the Hamiltonian of atoms A and B; $H_F$ is the free field Hamiltonian; and $H'_{eff}(A)$ and $H'_{eff}(B)$ are the effective Hamiltonian relative to atoms A and B, respectively. The~second-order energy shift in the polarizabilities, including only terms containing the position of both atoms (the other terms do not contribute to their interaction energy), is
\begin{equation}
\label{DispInter}
\Delta E_{AB} = \sum_{\bk \lambda \bkp \lambda '} \frac {\langle \vac \rvert H'_{eff}(A) \lvert 1_{\bk \lambda}1_{\bkp \lambda '}\rangle \langle 1_{\bk \lambda}1_{\bkp \lambda '}\rvert H'_{eff}(B) \lvert \vac \rangle }{-\hbar ( \wk +\wkp )} + (A \leftrightarrow B) .
\end{equation}

{We wish to point out that, although~our effective Hamiltonian is a second-order one in the atomic dipole moments (see Equation~(\ref{Transformation1})), in~the present case of two atoms we can take the two dipole moments ${\bm \mu}_A$ and ${\bm \mu}_B$ as independent expansion parameters, and~thus~(\ref{DispInter}) is a second-order quantity in ${\bm \mu}_A$ and in ${\bm \mu}_B$. Fourth-order terms in ${\bm \mu}_A$ and in ${\bm \mu}_B$, which would yield the fourth-order correction to the Lamb shift (not depending from the interatomic distance), would require a fourth-order effective Hamiltonian and~are here of course neglected, coherently with our approximations.}

Taking into account the form (\ref{EffHam2}) of the effective Hamiltonian, and~using (\ref{ElectricField}), we get
\begin{eqnarray}
\label{DispInter2}
\Delta E_{AB} &=& -\frac 1{4\hbar} \sum_{\bk \lambda \bkp \lambda '} \frac{\alpha_A(\wk )\alpha_B(\wk )}{\wk +\wkp}
\Big[ \bfkl (\bR_A)\cdot \bfklp (\bR_A) \bfkls (\bR_B) \cdot \bfklps (\bR_B)
\nonumber \\
&\ & \ + \bfkl (\bR_B)\cdot \bfklp (\bR_B) \bfkls (\bR_A) \cdot \bfklps (\bR_A)
\Big] ,
\end{eqnarray}
where $\alpha_A(\wk )$ and $\alpha_B(\wk )$ are, respectively, the dynamic polarizability of atoms A and~B.

For atoms in the free space and using the mode functions (\ref{FreeSpace}), we obtain
\begin{equation}
\label{DispInter3}
\Delta E_{AB} = -\frac {2\pi^2\hbar}{V^2} \sum_{\bk \lambda \bkp \lambda '} \frac{\alpha_A(\wk )\alpha_B(\wk )}{\wk +\wkp} \left( \ekl \cdot \eklp \right)^2 e^{i(\bk +\bkp )\cdot \bR} ,
\end{equation}
where $\bR = \bR_A - \bR_B$ is the distance between the two atoms or molecules
{(we have assumed real polarization unit vector). Explicit evaluation of (\ref{DispInter3}), using
$\sum_\lambda (\ekl)_i (\ekl)_j = \delta_{ij}-\hat{k}_i \hat{k}_j$ for the polarization sums and in the continuum limit $\sum_\bk \rightarrow (V/(2\pi )^3)\int d^3k$, yields the well-known dispersion interaction energy between two isotropic atoms as obtained through standard fourth-order perturbation theory, by~summing the contributions of the twelve relevant time-ordered Feynman diagrams} \cite{Casimir--Polder48,Craig-Thirunamachandran98,Passante18}. {It scales as $R^{-6}$ in the near zone (nonretarded regime: $R \ll c/\omega_0$, $\omega_0$ being a main transition angular frequency of the atom) and~as $R^{-7}$ in the far zone (retarded Casimir--Polder regime, $R \gg c/\omega_0$); thus, the~(attractive) force between the two atoms scales as $R^{-7}$ and $R^{-8}$ in the near and far zone, respectively. Typical values of the dispersion force between two hydrogen atoms are: $\sim$$10^{-30}$ N for $R=10^{-7}$ m (near zone) and $\sim$$10^{-37}$ N for $R=10^{-6}$ m (far zone).}
If the appropriate mode functions are used, Equation~(\ref{DispInter2}) can be also exploited to evaluate the dispersion interaction in the presence of generic metallic boundaries, for~example, a reflecting mirror~\cite{Power-Thirunamachandran82,Spagnolo-Passante06}.
{These results could in principle be extended to generic electrically polarizable bodies. For~systems of three or more macroscopic bodies, the~non-additivity of dispersion interactions should, however, be taken into account~\cite{Milonni94,Barcellona-Passante15}, and~effective Hamiltonians can be very useful in dealing with nonadditive interactions.}

\section{Conclusions}
\label{Conclusions}
In this paper, we have considered second-order effective Hamiltonians in nonrelativistic quantum electrodynamics,
{which are very useful to obtain several radiative processes through simpler calculations, compared to the usual evaluations based on the multipolar-coupling Hamiltonian. Examples are energy shifts and Casimir--Polder atom--surface interactions, as~well as two- and many-body dispersion interactions between atoms or molecules, in~both the nonretarded (near zone) and the retarded (far zone)} regime, even in the presence of reflecting boundary conditions. We have reviewed and analyzed different expressions of the effective Hamiltonian, according to the specific system considered and the field states, stressing their limit of validity. The~presence or absence of poles in the response function of the atoms has been carefully analyzed, showing that in the case of the photon vacuum the response function has no poles and it is a real function, while in the other cases it is the atomic dynamical polarizability, obeying the Kramers--Kr\"{o}nig dispersion relations. In~a forthcoming publication, we will address possible extensions of these results to macroscopic polarizable bodies and Casimir interactions between macroscopic dielectric bodies, stressing in particular their relevance for the dielectric model to be used for the evaluation of such interactions between macroscopic~bodies.

\vspace{6pt}

\section*{Acknowledgements}
The authors gratefully acknowledge financial support from the Julian Schwinger Foundation and MUR.

\end{document}